\documentclass[aps,prd,showpacs,nofootinbib,preprintnumbers,twocolumn]{revtex4-1}

\usepackage{bm}
\usepackage{latexsym}
\usepackage{natbib}
\usepackage{url}
\usepackage{dcolumn}
\usepackage{color}
\usepackage{amsfonts,amssymb,amsmath}
\usepackage{graphicx,epsfig}
\usepackage{psfrag}
\usepackage{subfigure}
\usepackage{natbib}
\begin{document}

\title{Cosmological wheel or the  arrow  of time: A classical versus quantum perspective of gravity}


\author{ Bal Krishna Yadav}
\email[]{balkrishnalko@gmail.com}

\author{Murli Manohar Verma}
\email[]{sunilmmv@yahoo.com}

\affiliation{Department of Physics, University of Lucknow, Lucknow 226 007, India}
\date{\today}

\begin{abstract}

It is shown that the structures in the universe can be interpreted to show a closed wheel of time, rather than a straight arrow. An analysis in $f(R)$          gravity model has been carried out to  show that  due to local observations  a small arc at any  given spacetime point  would invariably indicate an arrow of time from past to future, though on a quantum  scale it is not a linear flow but a closed loop, a fact that can be examined through future observations.

\end{abstract}
\maketitle
\section{\label{1}Introduction}
The cosmological arrow of time has been explained in cyclic universe without any dissipation in the presence of scalar field\cite{b1}. The cosmological arrow of time may be linked to the thermodynamic arrow by the second law of thermodynamics. The time asymmetry is also associated with dissipative fluid as Tolman introduced a viscous fluid to generate an arrow of time in cyclic cosmology.  Eddington once related  an arrow to the increase of entropy in isolated systems. There is an approach related with the entropy of a system in which time asymmetry might be a feature of a subsystem to which we belong and therefore time's arrow may be perspectival\cite{b2}.
It is also shown that the dark energy (positive cosmological constant) also supports the time asymmetry\cite{b3}. One of the most suitable candidates for dark energy is the cosmological constant $\Lambda$,\cite{b4} but we do not know  the precise origin of it, in addition to coincidence and fine tuning problems.  The other approaches to explain the dark energy include  modified gravity models \cite{b5},\cite{b6} and the modified matter models\cite{b7},\cite{b8}.  $f(R)$ dark energy models are the  simplest modified gravity models. The time asymmetry is shown in $f(R)$ gravity using the dissipation of the scalar field.  Since time asymmetry, and therefore ``time flow'' is a fundamental empirical fact of the universe,  it can lead  to  a suitable form of the $f(R)$ dark energy models.

In section II, using solutions of the field equations of $f(R) = \alpha R^2$ dark energy model \cite{b9} we find out the form of potential due to $f(R)$ term in the modified gravity model. In section III, we show a local arrow of time by taking the perturbations in a time dependent larger mass $M(t)$. Finally results are concluded in section IV.
\section{\label{2}Dynamics of effective $f(R)$ potential}

We  consider the limit of  the Newtonian gravity in which many $N$ non-relativistic particles move in a field of a large mass $M$.  Here we have an additional background potential due to the  $f(R)$  model. In the absence of matter, for   $f(R)=\alpha R^{2}$  type model (where $\alpha$ is a constant), the Ricci scalar as function of Hubble parameter $H$ is given by
\begin{eqnarray} R(H) = R_{0}\left(\frac{H}{H_{0}}\right)^{2},\label{a1}       \end{eqnarray}
where  $ R_{0}$   and  $H_{0}$     are constants. Using the expression of Ricci scalar

\begin{equation}\label{a26}
R=6(2H^{2}+\dot{H})
\end{equation}
in spatially flat universe and the above expression, we get the scale factor
\begin{eqnarray} a(t) = a_{0}\left[\frac{\chi-\left(\frac{H}{H_{0}}\right)^\frac{3}{2}}{\chi-1}\right]^\frac{-1}{3},\label{a2}         \end{eqnarray}
where $a_{0}$ is a constant and $\chi=\frac{R_{0}}{12 H^{2}_{0}}$,
whereas an inversion of this expression yields
\begin{equation}\label{a3}
  H = H_{0}\left[\chi-(\chi-1)\left(\frac{a}{a_{0}}\right)^{-3}\right]^{\frac{2}{3}}
\end{equation}

In equilibrium statistical mechanics, we have $N$-particle system which performs a finite motion and can be described by a microcanonical equilibrium distribution at some fixed energy.

Here, we consider the simplest situation, $N=1$, a one particle system for which the initial microcanonical distribution is well defined.

The contribution of $f(R)=\alpha R^2$ model in the acceleration of a test mass $m$ in the field of a larger mass $M\gg m$ is given by
\begin{equation}\label{a4}
  \ddot{a} = \frac{H_{0}\chi}{[\chi-(\chi-1)a^{3}_{0}a^{-3}]^{\frac{1}{3}}} + \frac{H_{0}(\chi-1)a^{3}_{0}}{[\chi a^{9}-(\chi-1)a^{3}_{0}a^{6}]^{\frac{1}{3}}}
\end{equation}
where $\ddot{a}$ represents the second derivative of distance w.r.t. $t$.
Potential due to the mass $M$ is

\begin{equation}\label{a5}
  \phi_{M}(a)= -\frac{GM}{a}
\end{equation}
and that due to $f(R)$ for lower powers of $a$ is given by
\begin{equation}\label{a6}
  \phi_{f}(a)=-\frac{(1-\chi)^\frac{1}{3}a^{2}_{0}H_{0}}{a}-\frac{2\chi H_{0}a^{2}}{3a_{0}(1-\chi)^\frac{1}{3}}
\end{equation}
If $L$ is the angular momentum of the test particle then total potential is given by
\begin{equation}\label{a7}
  \phi(a) = \frac{L^{2}}{2m^{2}a^{2}}-\frac{GM}{a} -\frac{ (1-\chi)^\frac{1}{3}a^{2}_{0}H_{0}}{a}-\frac{2\chi H_{0}a^{2}}{3a_{0}(1-\chi)^\frac{1}{3}}
\end{equation}
The conserved energy of the test particle is
\begin{equation}\label{a8}
E=\frac{\dot{a}^{2}}{2} + \phi(a)
\end{equation}
The motion of the test particle is influenced by the matter mass $M$ and the $f(R)$ gravity. Term $\frac{L^{2}}{2m^{2}a^{2}}$ in equation $(7)$ ensures a bounded motion of the test particle, for $L=0$ it will fall into the central mass. For simplicity we can replace it by a hard wall imposed at small distance $a_{0}$.

Let us consider the characteristic scale of energy is $\bar{E}$ and that of distance is $\bar{a}$ and write $\phi(a)=\bar{E}\phi(r)$ in terms of dimensionless $r=\frac{a}{\bar{a}}$ and $\phi(r)$,
Now potential $\phi(a)$ can be written as
\begin{equation}\label{a9}
  \phi(r) = -\frac{\alpha}{r} -\frac{\beta}{r}-\frac{\gamma}{2} r^2,      r\geq r_{0}
  \end{equation}
where $\alpha= G M$, $\beta=(1-\chi)^\frac{1}{3}a^{2}_{0}H_{0}$ and $\gamma=\frac{4\chi H_{0}}{3a_{0}(1-\chi)^\frac{1}{3}}$ and $r>r_{0}$ is the hard wall-condition.
We have
\begin{equation}\label{a25}
r_{0}=\frac{a_{0}}{\bar{a}}.
\end{equation}

The potential is maximal at,
\begin{equation}\label{a10}
r_{m}= \left(\frac{\alpha+\beta}{\gamma}\right)^{\frac{1}{3}}.
\end{equation}
and its maximum value is,

\begin{equation}\label{a20}
\phi(r_{m})= -\frac{3}{2}(\alpha+\beta)^{\frac{2}{3}}\gamma^{\frac{1}{3}}.
\end{equation}

 The condition for bounded and unbounded motion is given by $\phi(r_{m})>E>\phi(r_{0})$ and $\phi(r_{m})<E$ respectively. The maximal potential $\phi(r_{m})$ and maximal distance $r_{m}$ depends upon $\alpha$,$\beta$ and $\gamma$.

 We have
 \begin{equation}\label{a21}
 \gamma=\frac{4\chi H_{0}}{3a_{0}(1-\chi)^\frac{1}{3}}
 \end{equation}
  and $\gamma\rightarrow\infty$ when $(1-\chi)=0$ or $R_{0}=12H_{0}^{2}$.

In this case
 \begin{equation}\label{a22}
 \phi(r_{m})\rightarrow-\infty, r_{m}\rightarrow0.
\end{equation}
Therefore, for $ R_{0}=12H_{0}^{2}$ the system is unbounded forever.
\begin{figure}[h]
\centering  \begin{center} \end{center}
\includegraphics[width=0.50\textwidth,origin=c,angle=0]{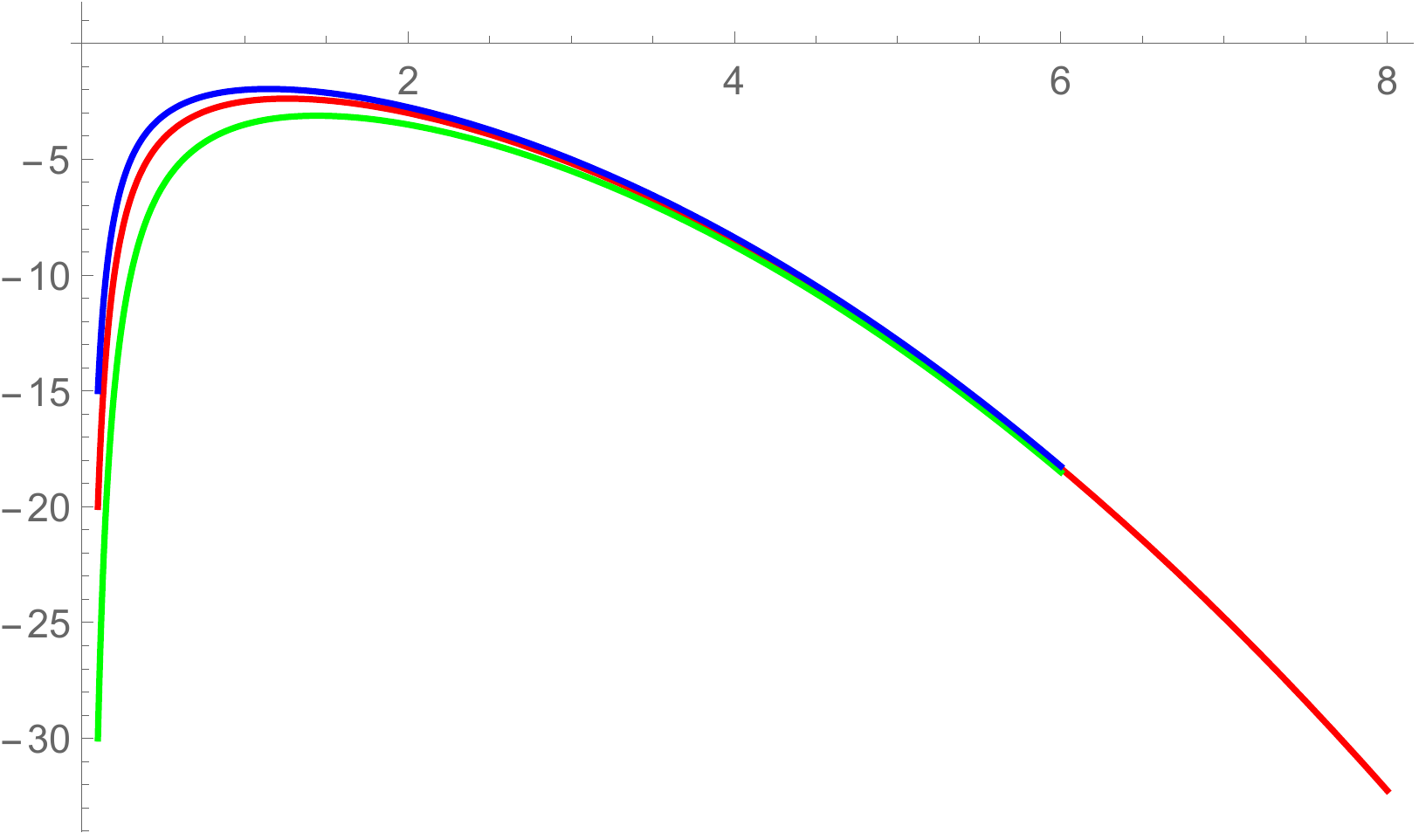}
\caption{\label{fig:p2} The effective potential $\phi(r)$ along Y-axis versus distance $r$ along X-axis. Red curve for $(\alpha=\beta=\gamma=1)$ and $r_{0}=0.10$. Blue curve for $\alpha=0.50,\beta=\gamma=1$ and $r_{0}=0.10$. Green curve $\alpha=2$ and$\beta=\gamma=1 $ and $r_{0}=0.10$ }\label{f2}

\end{figure}

\section{\label{3}Classical arrow but quantum  wheel of time}

We have a mechanical system executing a finite motion. Let us consider the situation when the large mass $M$ is time-dependent, if the  test particle’s motion is bounded and perturbation in mass $M(t)$  is slow, the energy $E(t)$   can be described by the adiabatic invariant phase-space volume\cite{b10}
\begin{equation}\label{a11}
    \int (da) (dP) \upsilon \left[E(t) - \frac{P^{2}}{2}-\phi(a,M(t))\right]
\end{equation}
where $P$ is the momentum and $\upsilon [x]$ is the step function. The entropy of a microcanonical equilibrium state is given as the logarithm of equation$(11)$. Its adiabatic conservation relates to the second law of thermodynamics.
 Integrating over $P$ in equation $(11)$ and writing in dimensionless quantities we get
\begin{equation}\label{a12}
  J =\int^{\bar{r}}_{r_{0}}dr \sqrt{\epsilon -\phi(r,\alpha(t))}
\end{equation}
where $\epsilon(t)= \frac{E(t)}{\bar{E}}$ and $\bar{r}$ is the maximal distance for the finite motion at time $t$.
For a given $\alpha(t)$, $\beta(t)$ and $\gamma(t)$, $\bar{r}(t)$ is always smaller than the largest possible distance of the bounded motion:

\begin{equation}\label{a13}
 \bar{r}(t)\leq r_{m}(t)= \left(\frac{\alpha+\beta}{\gamma}\right)^{\frac{1}{3}}
 \end{equation}

There are two cases in which bounded motion can turn to unbounded one.It depends upon the time dependent larger mass $M(t)$.

(1) When $M(t)\varpropto\alpha(t)$  decrease slowly. During the slow decrease of $\alpha(t)$, $\epsilon (t)$ grows faster than maximal potential energy $\phi (r_{m})$ . Here the change in $\alpha(t)$ is slow, otherwise, $\epsilon (t)$ will not change much and will stay bound. It is shown in Fig. $2$ where the  dashed curve represents the maximum value of potential energy $\phi(r_{m})$ for $\alpha=\beta=\gamma=1 $ and $r_{0}=0.10$.The curves below the dashed line refers to a bounded motion. It is clear that when $\alpha (t)$ decreases slowly, this motion becomes unbounded, i.e., its energy rises  above the dashed line.

\begin{figure}[h]
\centering  \begin{center} \end{center}
\includegraphics[width=0.50\textwidth,origin=c,angle=0]{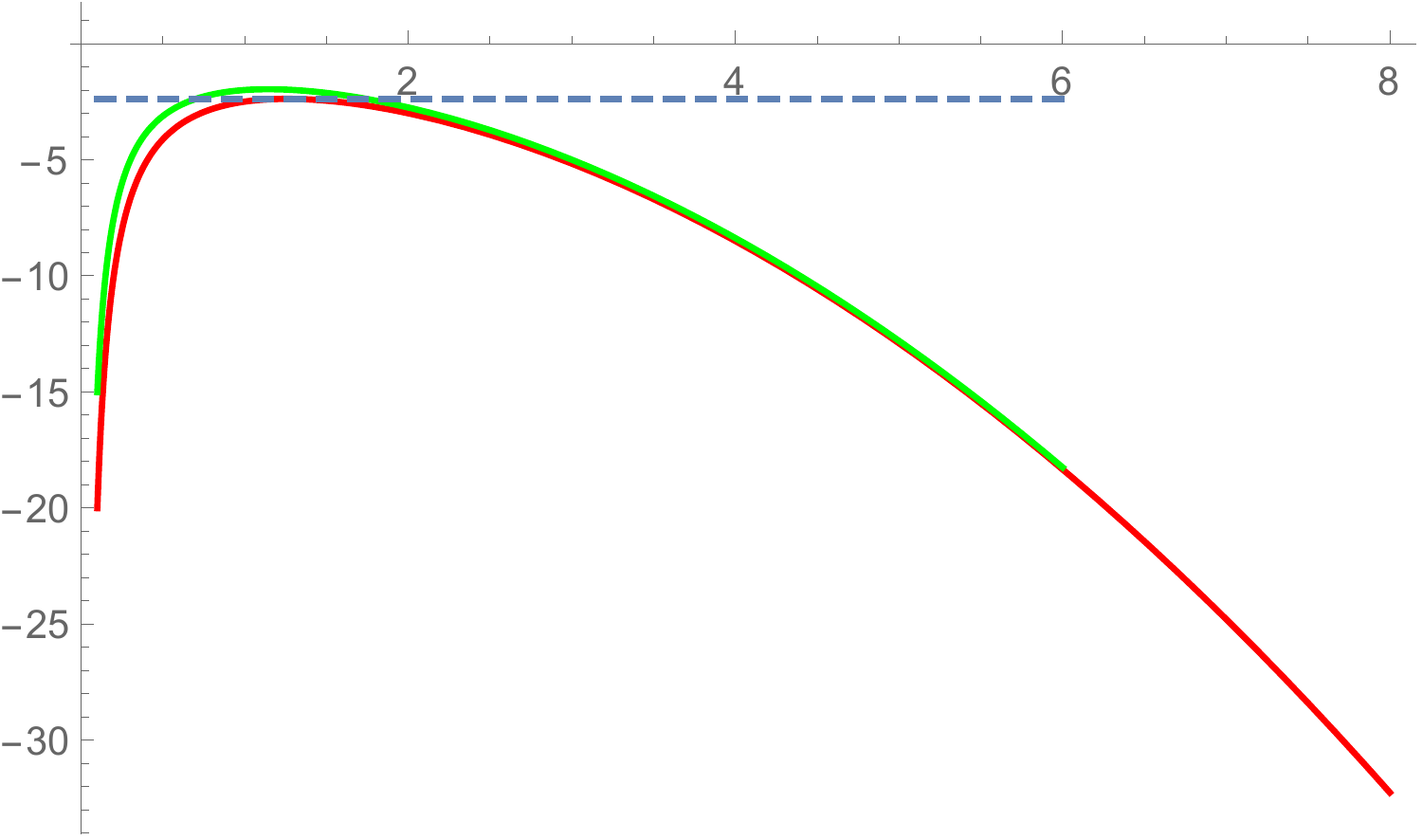}
\caption{\label{fig:p2} The effective potential $\phi(r)$ along Y-axis versus distance $r$ along X-axis. Red curve for $(\alpha=\beta=\gamma=1)$ and $r_{0}=0.10$. Green curve for $\alpha=0.50,\beta=\gamma=1$ and $r_{0}=0.10$. Dashed curve is $\phi(r_{m})$ for $\alpha=\beta=\gamma=1 $ and $r_{0}=0.10$. }\label{f3}

\end{figure}

(2) When  $M(t)\varpropto\alpha(t)$ increases. In this case the value of distance for maximal potential $r_{m}$ increases and the value of  energy $\epsilon(t)$ decreases. If  $\alpha(t)$         changes suddenly, then $\epsilon(t)$  will not change much and the motion will become unbounded\cite{b11}, it is shown in the Fig.$3$. This is related to the fact that the energy $\epsilon (t)$ will not change, since during a sudden change in $\alpha (t)$ the force changes by a finite amount in a small-interval. Therefore the acceleration changes by finite amount, while the changes in the coordinates and velocity are negligible.
\begin{figure}[h]
\centering  \begin{center} \end{center}
\includegraphics[width=0.50\textwidth,origin=c,angle=0]{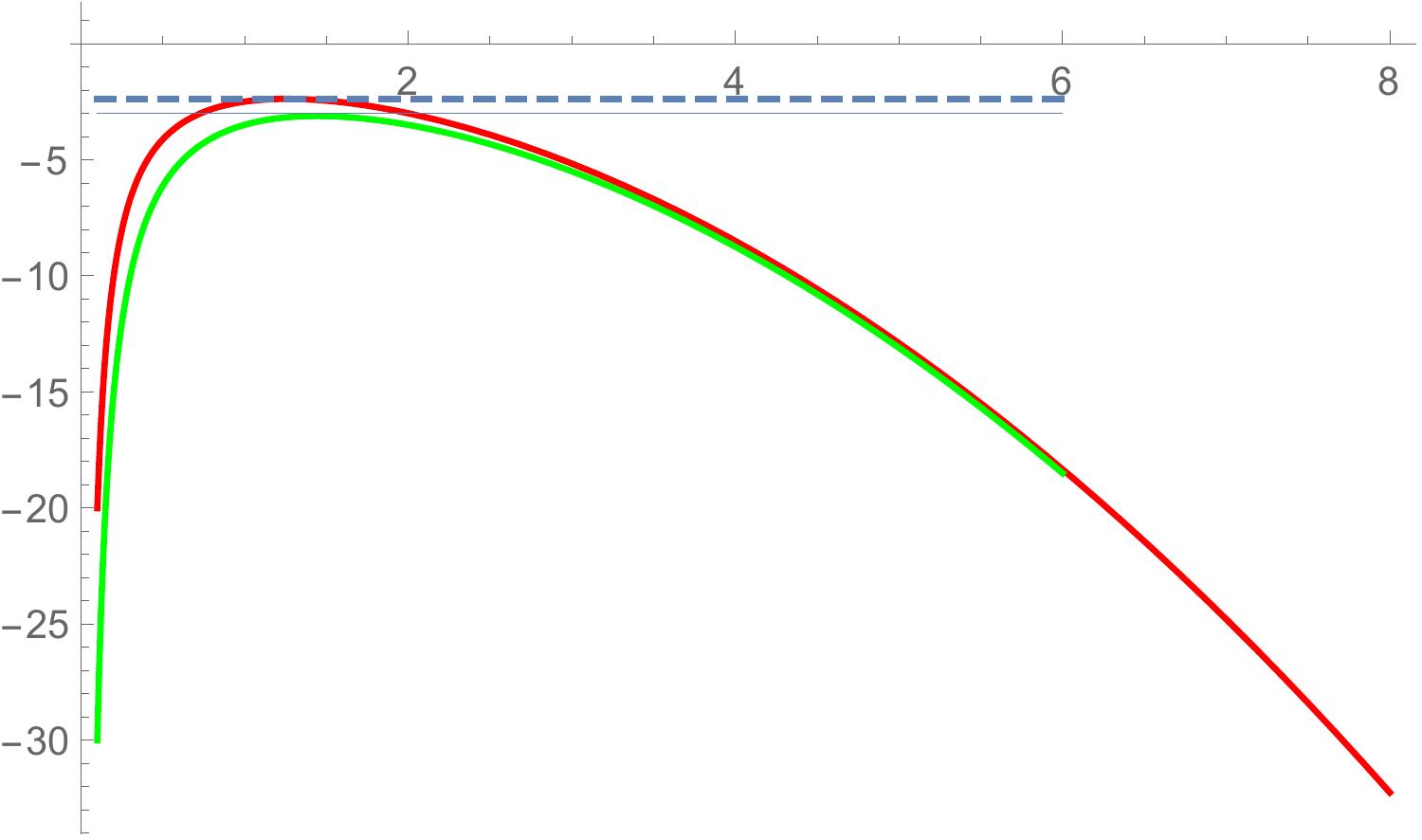}
\caption{\label{fig:p2} The effective potential $\phi(r)$ along Y-axis versus distance $r$ along X-axis. Red curve for $(\alpha=\beta=\gamma=1)$ and $r_{0}=0.10$. Green curve for $\alpha=2,\beta=\gamma=1$ and $r_{0}=0.10$. Dashed curve is $\phi(r_{m})=-2.38$, $\alpha=\beta=\gamma=1 $ and $r_{0}=0.10$ }.Thin line is an example of finite motion at energy $\epsilon =-3$. When $\alpha$ slowly changes from $\alpha=1$ to $\alpha=2$, the energy decreases and always refers to finite motion. If $\alpha$ changes sufficiently fast, the initial energy does not change much and now motion becomes unbounded.    \label{f4}

\end{figure}
Therefore, in both scenarios the motion changes from bounded to unbounded. This is an example of irreversibility of time.

The detailed structural analysis of cosmic matter distribution shows that the maximal Lyapunov exponent is given by
\begin{equation}\label{a13}
  \frac{1}{t}\ln \frac{\mod \delta z(t)}{\delta z_{0}}\rightarrow \sqrt{\lambda}
\end{equation}
for $t\rightarrow\infty$          . It determines the increasing phase space volume over small arc of time. However, the Lyapunov spectrum over the global cosmic scales leads to a closed Hamiltonian with constant phase volume, resulting in a constant entropy. This implies the existence of a closed loop of time perhaps as a wheel,  over the scales of Lyapunov time when the nearby trajectories are well resolved for the stability of the universe.

\section{\label{4}Conclusion}

We have argued that the large scale structures with the background  $f(R)$       contribution to potential must show a repulsive order causing the unbounded motion of the test particles such that the phase space volume remains constant over those scales. We found an asymmetry resulting  due to the unbound motion of the initially bound system. Owing to this fact, any local observation of structures would invariably  reveal a straight arrow of time just like a tangent to a small local arc of a circle or loop. However,  on the quantum  scales the future analysis is expected to show that the time axis must turn into a wheel and the usual arrow of time would have vanishing features.

\end{document}